\documentclass[fleqn,12pt,twoside]{article}
\usepackage{espcrc1}

\usepackage{epsfig}
\usepackage[figuresright]{rotating}

\title{High energy scattering in QCD as a statistical process
}

\author{S. Munier\thanks{Membre du CNRS, France. Permanent address: CPHT,
\'Ecole Polytechnique.}
\address{Centre de physique th\'eorique,
unit\'e mixte de recherche du CNRS 
(UMR 7644)\\ \'Ecole Polytechnique, 91128 Palaiseau cedex, France}
\address{Dipartimento di fisica, Universit\`a di Firenze,
50019 Sesto F., Florence, Italy}
}
       
\begin{document}

\maketitle

\begin{abstract}
The scattering of two hadronic objects at high energy is similar to
a reaction-diffusion process described by the stochastic Fisher-Kolmogorov
equation. 
This basic observation enables us to derive universal
properties of the scattering amplitudes in a straightforward way, 
by borrowing some general results from statistical physics.
\end{abstract}


\section{An event-by-event picture of high energy scattering}

We consider the scattering of a dipole of size
$r$ off a target $|q\bar q\rangle$-dipole of size $r_0$
as a toy model for high energy scattering.
In the QCD parton model and in the rest frame of the target,
the projectile interacts
through one of its quantum fluctuations 
$|q\bar q gggg\cdots\rangle$ which are produced by
QCD radiation. At high energy, these fluctuations 
are dominated by dense
gluonic states, generically called ``color glass
condensate'' \cite{VW}.

It proves useful to view the partonic configurations
as collections of color dipoles \cite{M} 
characterized by their transverse sizes $r_1,...,r_n$. 
Each of these dipoles may interact with the target according to
the elementary
amplitude $T_{el}(r_i,r_0)\sim \alpha_s^2 (r_<^2/r_>^2)$, 
where $r_<=\min(r_i,r_0)$ and $r_>=\max(r_i,r_0)$.
The total amplitude for a given Fock state realization reads 
\begin{equation}
T(r,r_0)=\sum_{i=1}^n T_{el}(r_i,r_0)\ .
\label{linear}
\end{equation}
Roughly speaking, $T$ is counting the number of dipoles
within a bin of size 1 centered around 
$\rho\equiv\log r_0^2/r^2$,
with a weight given by the interaction strength $\alpha_s^2$.

Let us describe the rapidity evolution of a given Fock state.
An increase of the rapidity of the projectile opens up the phase
space for each
dipole to split into two new dipoles
(this is the dipole interpretation of gluon branching).
The probability of such a splitting to occur is given 
by the Balitsky-Fadin-Kuraev-Lipatov 
(BFKL) kernel.
The typical partonic configurations that drive the high energy evolution
are those for which the newly created dipoles have
sizes of the order of the size of their parent dipole. This means that 
the parton density grows essentially 
diffusively under rapidity evolution.

The above picture is valid provided the interaction is not too strong, 
i.e. as long as $T\ll 1$. In the bins 
where the number of partons gets large and $T$ reaches~1,
Eq.~(\ref{linear}) and the rapidity evolution itself
are supplemented by nonlinear terms which tame the growth of $T$
in such a way that the unitarity
bound $T\leq 1$ be respected.
At that point, the dipole picture itself breaks down.
The exact mechanism
of how unitarization is realized 
(which involves gluon recombinations
and multiple scatterings) is still not fully understood, 
but such information is not required to get most of the properties
of the amplitude, see 
Eqs.~(\ref{satscale2},\ref{amp2}).

The picture of high energy scattering that 
we have just outlined is
essentially that of a reaction-diffusion 
process in a system made of
$N=1/\alpha_s^2$ particles \cite{IMM}.

So far, we have being focussing on one particular
Fock state realization, which corresponds to
one given event in an experiment. The amplitude $T$
is the scattering amplitude 
for that given partonic state,
which obviously is not an observable since partonic configurations
are random and cannot be selected experimentally. 
The physical amplitude is the average
of $T$ over all partonic configurations
accessible at rapidity~$Y$:
\begin{equation}
A(Y,r)=\langle T(r,r_0)\rangle_Y \ .
\label{physamp}
\end{equation}
In the following sections, we will study the 
properties of $A$ that can be deduced
from our reaction-diffusion picture.


\section{Mean field approximation and the FKPP equation}

As a first step,
we enforce a mean field approximation by neglecting 
the fluctuations of $T$ 
(which are due to statistical 
fluctuations in the dipole number), in which case $A=T$. 
The rapidity evolution
of $A$ is well-understood in this approximation: 
it is given by the Balitsky-Kovchegov (BK)
equation \cite{BK}, which was recently shown to belong to
the universality class of the Fisher-Kolmogorov-Petrovsky-Piscounov
(FKPP) equation \cite{MP}. The latter reads
\begin{equation}
\partial_{\bar\alpha Y} A=\partial_\rho^2 A + A - A^2 \ ,
\label{FKPP}
\end{equation}
where $\bar\alpha\equiv\alpha_s N_c/\pi$.
The first two terms in the right handside stand for the growing
diffusion of the partons, while the last nonlinear term tames
this growth so that $A$ complies with the unitarity limit.
While the exact mapping between the BK and the FKPP 
equations~(\ref{FKPP})
has not been exhibited, the universality class was
unambiguously identified~\cite{MP}.
From the properties of the solutions of equations
belonging to that class, it became clear that the two equations
have the same large-$Y$ asymptotics, 
up to the replacement of the few parameters 
that characterize the diffusive growth, and that
have to be taken from the BFKL kernel in the case of QCD.

\begin{figure}[htb]
\begin{minipage}[t]{77mm}
\epsfig{file=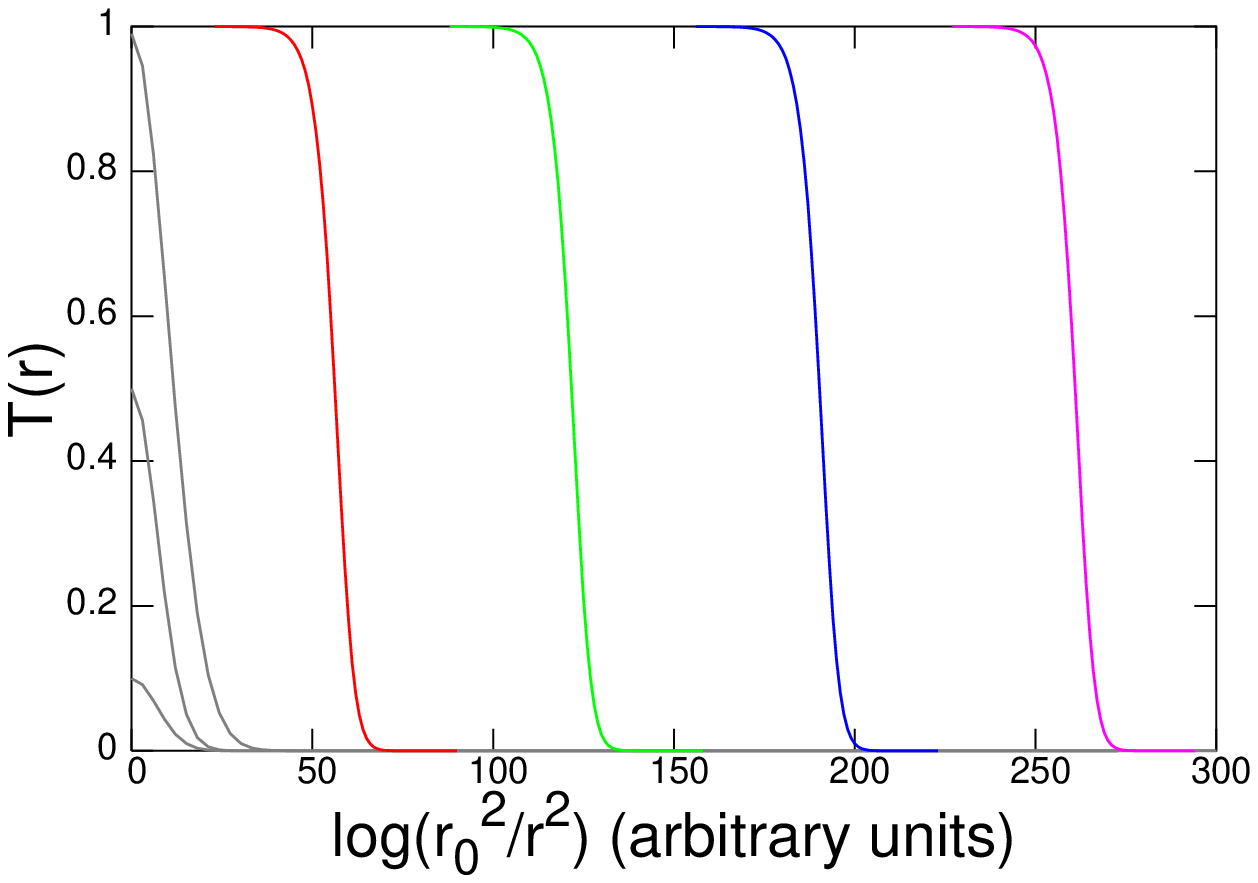,width=77mm}
\caption{\label{fig1}Picture of the evolution of $T(r)$ with rapidity
(from left to right).
Initially at $Y=0$, $T\sim T_{el}(r,r_0)$. When rapidity increases,
more dipoles populate the typical Fock state configuration, leading 
to a growing diffusion for $T$. 
Shortly after the unitarity limit $T=1$ is reached,
a traveling wave forms.}
\end{minipage}
\hspace{\fill}
\begin{minipage}[t]{77mm}
\epsfig{file=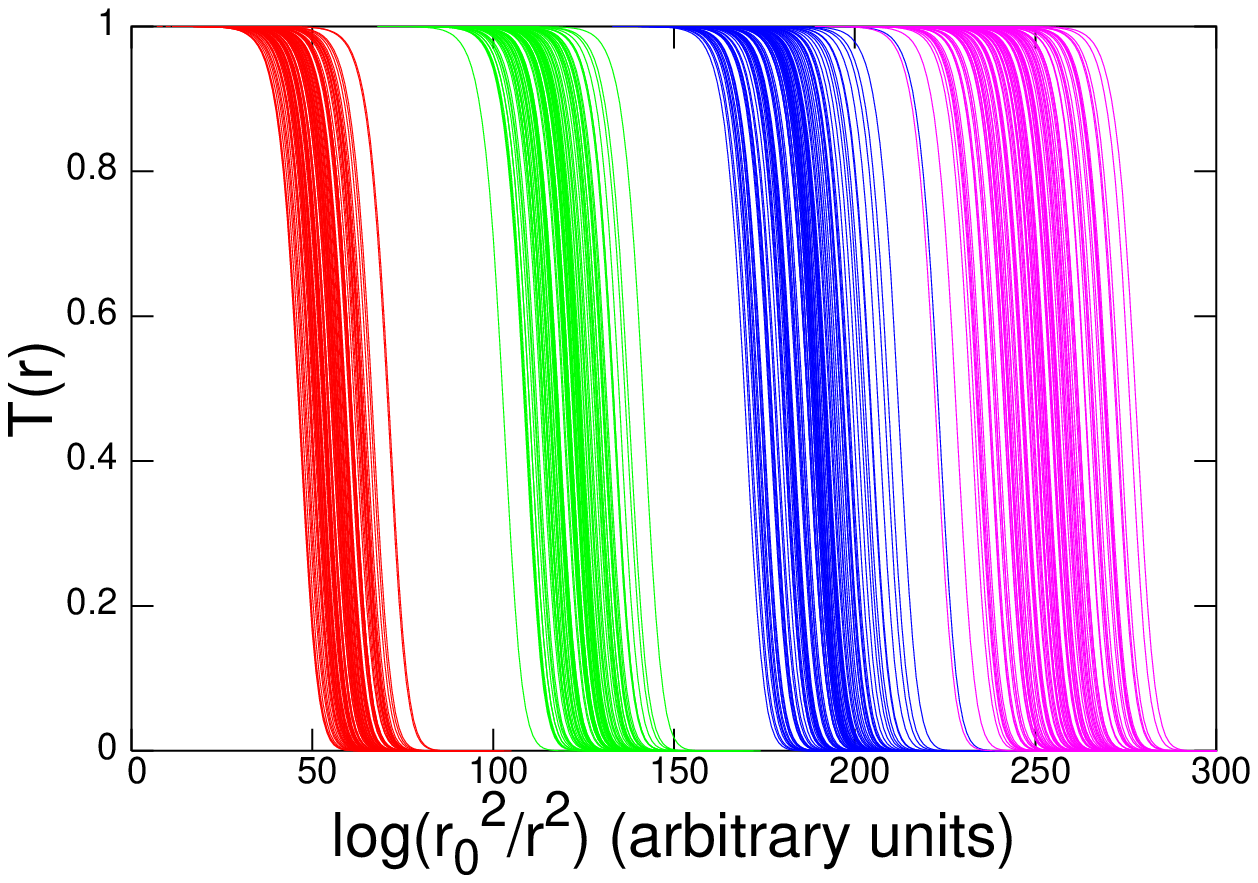,width=77mm}
\caption{\label{fig2}Different evolutions of a given 
initial condition over the
rapidity ranges $Y_0$, $2Y_0$, $3Y_0$, $4Y_0$ (the bunches of
curves gather 100 different realizations). 
One sees the dispersion of the front positions,
which is of diffusive nature and thus of order 
$\Delta(\ln r_0^2 Q_s^2(Y))\sim\sqrt{\bar\alpha Y}$.
}
\end{minipage}
\end{figure}

The FKPP equation admits asymptotic traveling waves (see Fig.~\ref{fig1}), 
namely at large rapidity, $A$ boils down to a function of 
a single variable
\begin{equation}
A(Y,r)=A(r^2 Q_s^2(Y))\ ,
\label{amp1}
\end{equation}
where the momentum $Q_s$ is called the saturation scale and
is defined, for example, 
by the requirement that $A(Y,1/Q_s(Y))$ be
some predefined number $A_0\sim{\cal O}(1)$.
The scaling law~(\ref{amp1}) is known as 
``geometric scaling'' \cite{SGK}.
The rapidity dependence of $Q_s$ was found to be \cite{GLR,MT,MP}
\begin{equation}
\ln r_0^2 Q_s^2(Y)=\bar\alpha\frac{\chi(\gamma_0)}{\gamma_0}Y\ ,
\ \ \chi(\gamma)=2\psi(1)-\psi(\gamma)-\psi(1-\gamma)\ ,
\ \ \frac{\chi(\gamma_0)}{\gamma_0}=\chi^\prime(\gamma_0)
\label{satscale1}
\end{equation}
where we have put back the parameters derived from
the BFKL kernel $\chi(\gamma)$.
The last equation in~(\ref{satscale1}) defines $\gamma_0$ \cite{GLR}.
Eq.~(\ref{satscale1}) captures the leading-$Y$ behavior:
two further 
subleading terms have been obtained recently \cite{MT,MP}. 
The asymptotic
form of the
amplitude is also known for $r>1/Q_s(Y)$ \cite{MT,MP}, 
as well as the first
corrections to the scaling~(\ref{amp1}) \cite{MP}.

A very important property is that $Q_s(Y)$ is completely
determined by the small-$A$ tail which drives the
evolution of the whole front. 

\section{Beyond the mean field: the ``high energy QCD/sFKPP'' correspondence}

The mean field description outlined above is justified
when the number of dipoles of a given size is large. 
However, in regions in which the
amplitude $T$ is of order $\alpha_s^2$, the statistical
fluctuations dominate and the mean field approximation breaks down. 
Both the fact that the number of
dipoles of a given size is finite, and
that this number fluctuates from event to event resulting in
fluctuations $\delta T\sim\alpha_s\sqrt{T}$,
are neglected in the mean field description~(\ref{FKPP}).
Since the tail $T\ll 1$ determines $Q_s$, 
we anticipate that these features have
dramatic consequences.
An evolution equation corrected for these effects may be written as
\begin{equation}
\partial_{\bar\alpha Y} T
=\partial_\rho^2 T + T - T^2 + \alpha_s\sqrt{T(1-T)}\,\eta\ ,
\label{sFKPP}
\end{equation}
where $\eta$ is a Gaussian white noise.
With respect to Eq.~(\ref{FKPP}), we have added a noise term
that accounts for the statistical fluctuations in the dipole number.
Eq.~(\ref{sFKPP}) is
the stochastic Fisher-Kolmogorov (sFKPP) equation.

Significant progress has been achieved in the last few years in
understanding the properties of its solutions \cite{BD}.
It has been realized that the discreteness of the number of dipoles
induces large corrections to the saturation scale, namely
\cite{BD,MS,IMM}
\begin{equation}
\ln r_0^2 Q_s^2(Y)=\bar\alpha\frac{\chi(\gamma_0)}{\gamma_0}Y
-\bar\alpha\frac{\pi^2\gamma_0\chi^{\prime\prime}(\gamma_0)}
{2\ln^2 1/\alpha_s^2}Y\ .
\label{satscale2}
\end{equation}
On the other hand, the fluctuations in the dipole number
result in a diffusive wandering of
the saturation scale: different events have saturation
scales which typically differ by ${\cal O}(\sqrt{\bar\alpha Y}$). 
This is illustrated in Fig.~\ref{fig2} for
a reaction-diffusion model
that shares the essential properties of the full QCD problem
(note that $\alpha_s=10^{-5}$, therefore statistical fluctuations in the
tail are not distinguishable on the plot).
After having
averaged over many events to get the physical amplitude~(\ref{physamp}),
the wandering of the saturation scale
causes a drastic breaking of geometric scaling.

The variance of the wandering of the front 
has recently been quantified \cite{BD}, and this result enables us
to derive the 
scaling law for $A$ \cite{IMM} that replaces geometric scaling~(\ref{amp1}):
\begin{equation}
A(Y,r)=A\left(\frac{\ln r^2 Q_s^2(Y)}
{\sqrt{\bar\alpha Y/\ln^3 1/\alpha_s^2}}\right)\ .
\label{amp2}
\end{equation}

Very recent work \cite{ITMSW} has aimed at
matching these ideas with the
so-called Balitsky-Jalilian-Marian-Iancu-McLerran-Weigert-Leonidov-Kovner 
(JIMWLK) formulation of 
the color glass condensate \cite{VW}, 
at the expense of a
modification of the latter. 
We should stress
that in any case, 
the results~(\ref{satscale2},\ref{amp2}) 
directly stem from the physics of the
parton model, and are presumably exact solutions of QCD:
they are the leading terms in a large-$Y$ and small-$\alpha_s$ expansion.


\end{document}